\documentclass[apj]{emulateapj}

\usepackage{graphicx,amssymb,amsmath,times}

\newcommand{\lsi}{LS~I~+61$^{\circ}$303}

\begin{document}
\title{Detection of the $\gamma$-ray binary \lsi\ in a low flux state at Very High Energy $\gamma$-rays
with the MAGIC Telescopes in 2009} \shorttitle{Low flux state VHE
$\gamma$-rays emission from \lsi\ } \shortauthors{J. Aleksi\'c
et~al.}
%
\author{
J.~Aleksi\'c\altaffilmark{1}, E.~A.~Alvarez\altaffilmark{2},
L.~A.~Antonelli\altaffilmark{3}, P.~Antoranz\altaffilmark{4},
M.~Asensio\altaffilmark{2}, M.~Backes\altaffilmark{5},
J.~A.~Barrio\altaffilmark{2}, D.~Bastieri\altaffilmark{6},
J.~Becerra Gonz\'alez\altaffilmark{7,8}
% ...
}\altaffiltext{1} {IFAE, Edifici Cn.,
Campus UAB, E-08193 Bellaterra, Spain} \altaffiltext{2}
{Universidad Complutense, E-28040 Madrid, Spain} \altaffiltext{3}
{INAF National Institute for Astrophysics, I-00136 Rome, Italy}
\altaffiltext{4}{ Universit\`a  di Siena, and INFN Pisa, I-53100
Siena, Italy} \altaffiltext{5} {Technische Universit\"at Dortmund,
D-44221 Dortmund, Germany} \altaffiltext{6} {Universit\`a di
Padova and INFN, I-35131 Padova, Italy} \altaffiltext{7} {Inst. de
Astrof\'{\i}sica de Canarias, E-38200 La Laguna, Tenerife, Spain}
\altaffiltext{8} {Depto. de Astrof\'{\i}sica, Universidad de La
Laguna, E-38206 La Laguna, Spain}
% ...

\begin{abstract}
We present very high energy (VHE, E $>$ 100 GeV) $\gamma$-ray
observations of the $\gamma$-ray binary system \lsi~obtained with
the MAGIC stereo system between 2009 October and 2010 January. We
detect a 6.3$\sigma$ $\gamma$-ray signal above 400 GeV in the
combined data set. The integral flux above an energy of 300 GeV is
$F(E>300\mathrm{GeV})=(1.4 \pm 0.3_\text{stat} \pm
0.4_\text{syst}) \times 10^{-12} \mathrm{cm}^{-2}
\mathrm{s}^{-1}$, which corresponds to about 1.3\% of the Crab
Nebula flux in the same energy range. The orbit-averaged flux of
\lsi\ in the orbital phase interval 0.6--0.7, where a maximum of
the TeV flux is expected, is lower by almost an order of magnitude
compared to our previous measurements between 2005 September and
2008 January. This provides evidence for a new low flux state in
\lsi. We find that the change to the low flux state cannot be
solely explained by an increase of photon-photon absorption around
the compact star.

\end{abstract}


\section{Introduction}

The \lsi\ system consists of a Be star and a compact object of
still uncertain nature, either a neutron star or a black hole. Its
orbital period, which is most precisely measured in radio, is
$26.4960\pm0.0028$ days~\citep{Gregory:2002}. Soft X-ray outbursts
modulated with the same period as in the radio waveband were
reported by~\cite{Paredes_1997A&A...320L..25P} and changes in the
orbital evolution have been recently
studied~\citep{Torres_2010ApJ...719L.104T}. Many other orbital
parameters of the system are less precisely known and different
solutions have been proposed (see
\citealt{Casares:2005wn,Grundstrom:2006,Argona:2009}) but
observations indicate a highly eccentric orbit ($e=0.55\pm0.05$)
with the periastron passage at orbital phase
$\phi_{\mathrm{per}}=0.275$~\citep{Argona:2009}. These orbital
parameters are important for modelling the VHE emission of the
system as shown in,e.g.,
\cite{Sierpowska-Bartosik:2009ApJ...693.1462S} or
\cite{Dubus_rel_boosting_2010A&A...516A..18D}.

In 2006 the MAGIC collaboration discovered variable VHE
$\gamma$-ray emission from \lsi~\citep{MAGIC_lsi_science:2006vk}.
A following  extensive observational campaign in Fall 2006 found a
period for the VHE emission of $26.6\pm0.2$
days~\citep{MAGIC_lsi_periodic:2009ApJ...693..303A}. The VHE
$\gamma$-ray emission shows an outburst in the orbital phase
interval 0.6--0.7 with no significant $\gamma$-ray emission
detected during the rest of the orbit. In particular, no VHE
$\gamma$-ray signal was detected by MAGIC around the periastron
passage of the system. The data from Fall 2006 also suggested a
correlation between the X-ray and VHE $\gamma$-ray
flux~\citep{MAGIC_2008ApJ...684.1351A}. An extensive
multi-wavelength campaign conducted in 2007, including MAGIC,
XMM-Newton and \emph{Swift}, provided strong evidence for the
X-ray/VHE $\gamma$-ray flux correlation in strictly simultaneous
data~\citep{MAGIC_lsi_xrayvhe:2009ApJ...706L..27A}. In contrast,
no correlation was found between the radio wavelength flux and the
VHE $\gamma$-ray flux from the Fall of 2006
campaign~\citep{MAGIC_2008ApJ...684.1351A}.

The VHE emission of \lsi\ was confirmed by VERITAS observations
between 2006 Sep and 2007
Feb~\citep{Veritas_lsi_discovery:2008ApJ...679.1427A}. However, in
observations conducted by the VERITAS collaboration in Fall 2008
and early 2009, no VHE signal was detected. More recent VERITAS
observations in Fall 2009 (the same time period as in the present
paper) also yielded only upper limits  for VHE emission from
\lsi~\citep{VERITAS_lsi_2011arXiv1105.0449A}. Very recently the
VERITAS collaboration reported a detection of the system with a
significance of more than
$5\sigma$~(\citealt{VERITAS_lsi_2011arXiv1105.0449A}) between
orbital phases 0.05 and 0.23. This places the detection at
superior conjunction and 5.8 to 1.3 days before the periastron
passage. No VHE $\gamma$-ray emission was previously detected in
this phase range.

The binary system was observed in high energy (HE, $0.1-
100\mathrm{ GeV}$) $\gamma$-rays by
EGRET~\citep{EGRET_disc_1997ApJ...486..126K}  but the large
position uncertainty of the source and inconclusive variability
studies of the emission, prevented its unambiguous identification.
The positional association with \lsi\ was only achieved following
HE $\gamma$-ray observations by
AGILE~\citep{AGILE_discovery_2009A&A...506.1563P}. More recently,
\emph{Fermi}/LAT found that the HE $\gamma$-rays are periodically
modulated in very good agreement with the (radio) orbital
period~\citep{FERMI_disvoery_2009ApJ...701L.123A} establishing
beyond doubt that the signal origins from \lsi. The HE outburst
was not, however, observed at the same phases as the VHE outburst
but occurred between phase 0.3 and 0.45 just after the periastron
passage. This difference in phase may indicate that different
processes are responsible for the HE and VHE $\gamma$-ray
emission. On the other hand, the same process might produce both
emissions if the GeV $\gamma$-rays are produced by inverse Compton
(IC) pairs cascading developing in the radiation field of the
star. This cascade would reduce the TeV emission and enhance the
GeV emission when the compact object is close to the star. For
more details on such a scenario see~\cite{2006MNRAS.368..579B}.
Another possibility is that the shift in the peak emission could
be caused by a different location of the $\gamma$-ray production
site in the system~\citep{2011A&A...527A...9Z}. We note that no
simultaneous VHE observations are available at the same epoch (Aug
2008 to Jan 2009) of the first reported Fermi observations
\citep{FERMI_disvoery_2009ApJ...701L.123A}. An unambiguous
interpretation of the non simultaneous SED from MeV to TeV
energies of \lsi\ is not possible because the system might have
changed its VHE emission in the meantime.

Two principal scenarios have been proposed to explain the
non-thermal emission from \lsi: an accretion powered microquasar
(e.g.
\citealt{Romero_2005ApJ...632.1093R,Bednarek:2006,Gupta_2006ApJ...650L.123G,Bosch-Ramon_2006A&A...459L..25B})
and a rotation-powered compact pulsar wind (e.g.
\citealt{Dubus:2006,Sierpowska-Bartosik:2009ApJ...693.1462S,Zdziarski_2010MNRAS.403.1873Z}).
An alternative model assumes that the compact object is an
accreting magnetar and that the  $\gamma$-rays are produced along
the accretion flow onto the magnetar~\citep{2009MNRAS.397.1420B}.
High resolution radio measurements
\citep{Dhawan_2006smqw.confE..52D} show an extended structure
varying in shape and position as a function of the orbital phase.
While this was taken as evidence for a pulsar wind interacting
with that of the Be star other interpretations were suggested as
well~\citep{Romero_2007}. Neither of the two proposed scenarios
could be validated by accretion disk features, e.g. a thermal
component in the X-ray spectrum, or the presence of pulsed
emission at any wavelength. Thus the engine behind the VHE
emission remains an open question.

Here we present new observations of \lsi\ conducted with the MAGIC
stereo system. This has twice the sensitivity of the previous
MAGIC campaigns, and results in a significant detection of the
binary system during a newly identified low flux state. We briefly
discuss the observational technique and the data analysis
procedure, present the VHE $\gamma$-ray light curve of the source,
and put the results in context of the previous VHE $\gamma$-ray
observations of this system.

\section{Observations}

The observations were performed between 2009 Oct 15 and 2010 Jan
22 using the MAGIC telescopes on the Canary island of La Palma
($28.75^\circ$N, $17.86^\circ$W, 2225~m a.s.l.), from where \lsi\
is observable at zenith distances above 32$^{\circ}$. The MAGIC
stereo system consists of two imaging air Cherenkov telescopes,
each with a 17~m diameter mirror. The observations were carried
out in stereo mode, meaning only shower images which trigger
simultaneously both telescopes are recorded. The stereoscopic
observations provide a $5\sigma$ signal above 300~GeV from a
source which exhibits 0.8\% of the Crab Nebula flux in 50 hours
observation time, a factor of two more sensitive than our single
telescope campaign on \lsi\ in 2007. Further details on the design
and performance of the MAGIC stereo system can be found
in~\cite{MAGIC_stereo_performance}.

The \lsi\ data set spans four orbits of the system, with two
observed for only one and three nights, respectively. The data
taken in 2009 Oct and 2009 Nov were restricted to moonless nights.
The data sample the orbital phases 0.55 to 0.98 for 2009 Oct , and
0.58 to 1.02 for 2009 Nov , the last night of which is in the next
orbital cycle. The data recorded in 2010 Jan  cover the phases
0.22 to 0.32 and were obtained during moonlight conditions (see
Table~\ref{tab:tab1}). All data were taken at zenith angles
between 32$^{\circ}$ and 48$^{\circ}$. After pre-selection of good
quality data a total of 48.4~hours of data remained for the
analysis. The observation strategy aimed to cover consecutive
nights with at least three hours of observation in each individual
night. Due to adverse observation conditions such as bad weather,
the data set does not have uniform coverage during the orbital
phases and some nights have shorter observation times than the
planned three hours.

\section{Data Analysis}

The data analysis was performed with the standard MAGIC
reconstruction software. The recorded shower images were
calibrated, cleaned and used to calculated image parameters
individually for each telescope. The energy of each event was then
estimated using look up tables generated by Monte Carlo (MC)
simulated $\gamma$-ray events. The events that simultaneously
triggered both telescopes (the so-called stereo events) were then
selected\footnote{This step is only needed for the 2009 October
data where no hardware stereo trigger was yet available.} and
further parameters, e.g. the height of the shower maximum and the
impact parameter from each telescope, were calculated. The gamma
hadron classifications and reconstructions of the incoming
direction of the primary shower particles were then  performed
using the Random Forest (RF) method~\citep{magic:RF}. The RF
calculates a variable called hadronness which is a measure of the
probability that an event is of hadronic origin. Finally, the
signal selection used cuts in the hadronness (calculated by the
RF) and in the squared angular distance between the shower
pointing direction and the source position ($\theta^2$). The
energy dependent cut values were determined by optimizing them on
a sample of events recorded from the Crab Nebula under the same
zenith angle range and similar epochs than the \lsi\ data. For the
energy spectrum and flux, the effective detector area was
estimated by applying the same cuts used on the data sample to a
sample of MC simulated $\gamma$-rays. Finally, the energy spectrum
was unfolded, accounting for the energy resolution and possible
energy reconstruction bias~\citep{magic:unfolding}.

In this analysis we use for the estimation of the detection
significance a set of cuts optimized to yield the highest
significance on a sample of Crab Nebula data under similar
observation conditions as the \lsi\ data set. These cuts are then
applied to a set of simulated MC $\gamma$-rays to estimate the
energy threshold of the detection plot ($E_{\mathrm{th}}=400
\mathrm{GeV}$). For the light curve and spectrum determination
softer cuts are used to reduce systematic effects and provide a
lower energy threshold by sacrificing the highest significance.

\section{Results}

\begin{figure}[tbp]
  \centering
  \includegraphics[width=\linewidth]{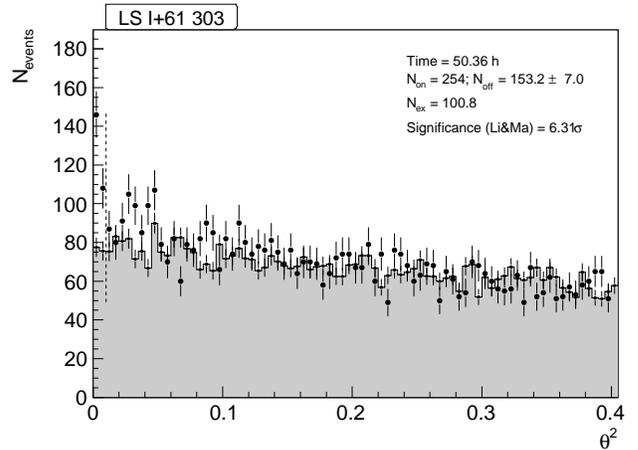}
  \caption{The squared angular distance between pointing
direction of the shower and the source position  ($\theta^2$-plot)
for the position of \lsi\ (points) and the simultaneous determined
background regions (grey shaded histogram) for the total 2009/2010
MAGIC data set. $\mathrm{N}_{\mathrm{on}}$ is the number of events
at the source position, $\mathrm{N}_{\mathrm{off}}$ is the number
of background events, $\mathrm{N}_{\mathrm{ex}}$ is the number of
excess events
($\mathrm{N}_{\mathrm{ex}}=\mathrm{N}_{\mathrm{on}}-\mathrm{N}_{\mathrm{off}}$)
and the significance was calculated according to \cite{Li:1983fv}.
  }
  \label{fig:detection}
\end{figure}

The integral data set of 48.8 hours presented here results in a $6.3 \sigma$ detection
of VHE $\gamma$-ray emission above 400 GeV from \lsi\ (see
Fig.~\ref{fig:detection}). The integrated flux above 300 GeV is
\begin{equation*}
F(E>300 \mathrm{GeV}) =  (1.4 \pm 0.3_\text{stat} \pm
0.4_\text{syst}) \times 10^{-12} \mathrm{cm}^{-2} \mathrm{s}^{-1}.
\end{equation*}
corresponding to about 1.3\% of the Crab Nebula flux in the same energy range.

\subsection{Light curve}

\begin{figure}[tbp]
  \centering
  \includegraphics[width=\linewidth]{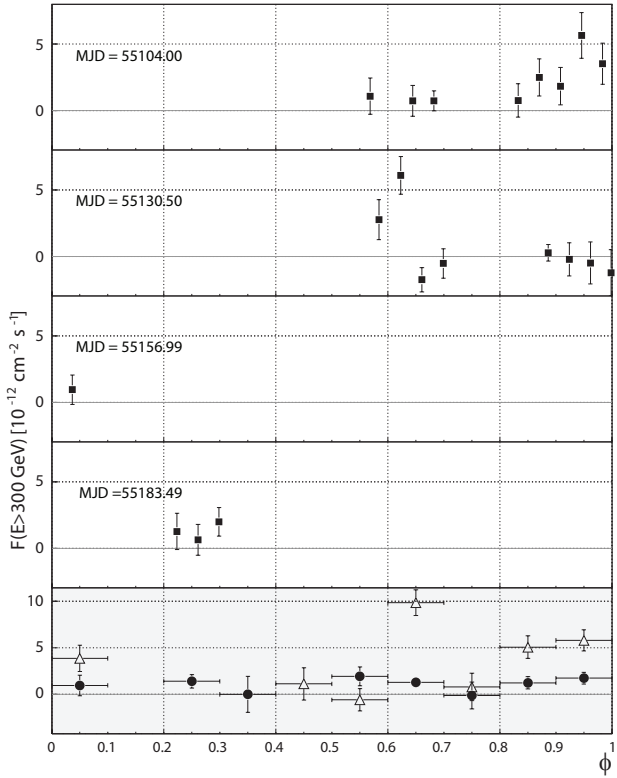}
  \caption{VHE ($E>300\text{GeV}$) $\gamma$-ray flux of \lsi\ as a
    function of the orbital phase for the four observed orbital cycles
    (four upper panels) and averaged for the entire observation time
    (lowermost panel, black points). The starting MJD of each orbital cycle is given in the
    corresponding panel. In the lower most panel we show as
    hollow triangles the previous published \citep{MAGIC_lsi_xrayvhe:2009ApJ...706L..27A}
    averaged fluxes per phase bin. Vertical error bars show $1\sigma$ statistical
    errors.}
  \label{fig:lc}
\end{figure}

We derived a nightly light curve above an energy of 300~GeV that
is shown in Fig.~\ref{fig:lc}. The measured fluxes and upper
limits are quoted in Table~\ref{tab:tab1}. A constant flux fit to
the light curve yields a $\chi^2 / \mathrm{dof}=42.15 / 19 $
($p=1.5\times 10^{-3}$) and hence is unlikely. Thus, as in
previous observations, the emission is variable and reaches a
maximum flux around orbital phase 0.62 of $F(E>300 \mathrm{GeV}) =
(6.1 \pm1.4_\text{stat} \pm 2.4_\text{syst}) \times 10^{-12}
\mathrm{cm}^{-2} \mathrm{s}^{-1}$, corresponding to 5.4\% of the
Crab Nebula flux. This is a much lower peak emission than detected
in our previous campaigns at the same orbital phases and sampled
with very similar cadence. For a more quantitative comparison of
the 2009 emission level with the previous MAGIC observations, we
included in Figure~~\ref{fig:lc} the lightcurve of the 2007 data
averaged in 0.1 phase bins.

\begin{table}
  \caption{Observation time, orbital phase, integral
flux (above
    300~GeV). Flux upper limit at the 95\% confidence level are quoted in
    case flux significance is $\lesssim 2 \sigma$
    ~\citep{Rolke:2004mj}. All errors are
    statistical only we estimate an additional systematic
    uncertainty of about 40\%. The systematic uncertainty is only important in case of
    comparing between different experiments.}
\vspace*{-0.3cm}
\begin{center}
\resizebox{8.5cm}{!}{\begin{tabular}{lccccc}
 \hline \hline
  Middle Time    & Obs. Time & Phase& Significance & Flux & Upper limit \\
      (MJD)   & (min)     &          &(pre-trial)  & 10$^{-12}$          & 10$^{-12}$    \\
      &           &            &  & (cm$^{-2}$ s$^{-1}$)& (cm$^{-2}$ s$^{-1}$)  \\
\hline
55119.07 & 138 & 0.57 & 0.8 & 1.1$\pm$1.4 & 4.0\\
55121.08 & 176 & 0.64 & 0.6 &0.7$\pm$1.2 & 3.2\\
55122.08 & 194 & 0.68 & 1.0 & 0.7$\pm$0.8 & 2.4\\
55126.06 & 104 & 0.83 & 0.6 &0.8$\pm$1.3 & 3.6\\
55127.06 & 137 & 0.87 & 1.9 &2.5$\pm$1.4 & 5.5\\
55128.07 & 137 & 0.91 & 1.4 &1.8$\pm$1.4 & 4.9\\
55129.06 & 132 & 0.95 & 3.8 &5.7$\pm$1.7 & \nodata\\
55130.06 & 135 & 0.98 & 2.5 &3.5$\pm$1.5 & \nodata\\
\hline
55145.97 & 140 & 0.58 & 2.0 &2.8$\pm$1.5 & \nodata\\
55147.01 & 221 & 0.62 & 4.9 &6.1$\pm$1.4 & \nodata\\
55148.00 & 216 & 0.66 & -1.8 &-1.8$\pm$0.9 & 1.0\\
55149.02 & 171 & 0.70 & -0.5 &-0.5$\pm$1.1 & 1.9\\
55153.99 & 123 & 0.89 & 0.5 &0.3$\pm$0.6 & 1.8\\
55154.98 & 149 & 0.92 & -0.2 &-0.2$\pm$1.2 & 2.5\\
55155.98 & 115 & 0.96 & -0.3 &-0.5$\pm$1.6 & 2.9\\
55156.95 & 73 & 1.00 & -0.7&-1.2$\pm$1.7 & 2.9\\
\hline
55157.97 & 82 & 0.04 & 0.9 &1.0$\pm$1.1 & 3.6\\
\hline
55215.90 & 134 & 0.22 & 1.0 &1.3$\pm$1.4 & 4.2\\
55216.90 & 161 & 0.26 & 0.6 &0.6$\pm$1.2 & 3.2\\
55217.90 & 165 & 0.30 &2.0 &2.0$\pm$1.1 & \nodata\\
\hline\hline
\end{tabular}}
\end{center}
\label{tab:tab1}
\end{table}

We found that the averaged emission level is dramatically lower
than measured in our campaigns from 2005 to
2007~\citep{MAGIC_lsi_science:2006vk,MAGIC_lsi_periodic:2009ApJ...693..303A,MAGIC_lsi_xrayvhe:2009ApJ...706L..27A}.
Not only had the flux changed but a VHE $\gamma$-ray excess was
also observed at phases other than those of the periodic outburst
between 0.6--0.7. The highest flux is, however, again detected in
an outburst during the interval 0.6--0.7, and the measurements in
the orbital cycle of 2009 Nov show the same burst profile as in
previous observations but with a reduced flux level. Whether the
outburst recurrence is still a periodic property of the VHE
emission for \lsi\, and whether it shows the same shape as in
previous campaigns, cannot be determined with the small number of
orbital cycles observed in this campaign.  It is noteworthy that
the outburst was not detected during the orbit observed in 2009
Oct.

The mean flux for all phase bins is given in Table~\ref{tab:tab2}.
The rather low mean values, even in the phase bin 0.6--0.7, of the
individual night peak emission indicates that most of the emission
of the system is contributed by only few nights instead of a
constant flux. A fit to a constant flux in the phase bin 0.6--0.7
yields a $\chi^2/ \mathrm{dof}=22.4 / 4$ being strongly
disfavored.

\begin{table}[tbp]
\caption{Average flux level above 300~GeV for each orbital
0.1--phase
    bin. Flux upper limit at the 95\% confidence level are quoted in
    case flux significance is $\lesssim 2 \sigma$.
    ~\citep{Rolke:2004mj}. All errors are
    statistical only we estimate an additional systematic
    uncertainty of about 40\%. The systematic uncertainty is only important in case of
    comparing between different experiments.}
\vspace*{-0.3cm}
\begin{center}
\begin{tabular}{cccc}
\hline \hline
 Phase bin & Flux & Flux upper limit \\
           & (10$^{-12}$cm$^{-2}$ s$^{-1}$)& (10$^{-12}$cm$^{-2}$ s$^{-1}$)\\
\hline
0.0--0.1         & 1.0  $\pm$ 1.1  & 1.3\\
0.1--0.2         & \nodata & \nodata \\
0.2--0.3         & 1.4 $\pm$ 0.7  & \nodata\\
0.3--0.4         & 0.0 $\pm$ 1.9  & 0.0\\
0.4--0.5         & \nodata  & \nodata  \\
0.5--0.6         & 1.9  $\pm$ 1.0 & 1.3 \\
0.6--0.7         & 1.3  $\pm$ 0.5 & \nodata  \\
0.7--0.8         & $-$0.1  $\pm$ 1.4 & 2.0 \\
0.8--0.9         & 1.2  $\pm$ 0.7  & 1.1 \\
0.9--1.0         & 1.7 $\pm$ 0.6   & \nodata\\
\hline\hline
\end{tabular}
\end{center}
\label{tab:tab2}
\end{table}

\subsection{Spectrum}

\begin{figure}[tbp]
  \centering
  \includegraphics[width=\linewidth]{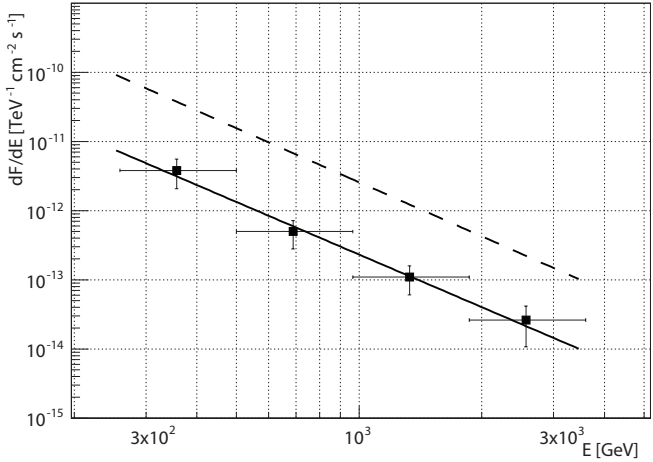}
  \caption{The spectrum of the complete \lsi\ data sample recorded by the MAGIC stereo system in 2009 is shown in
  black. The vertical errors are $1\sigma$ statistical errors.
  The fit to the most precise measured spectrum from Fall 2006~\citep{MAGIC_lsi_periodic:2009ApJ...693..303A} is shown
  as the dashed line. The 2009 spectrum is fitted by a simple
  power law as well and the fit parameters are compatible to our previous measurements from 2005 to 2008 (see text for discussion).
}
  \label{fig:spectra}
\end{figure}

The emission level of \lsi\ was too low during most phases to
obtain statistical significant phase dependent spectra. The total
signal, however, was sufficient to form a phase integrated
spectrum with good enough statistic per bin to perform a
Chi-square test, shown in Fig.~\ref{fig:spectra}.

The spectrum is well described by a simple power law
\begin{equation*}
\frac{\mathrm{d}F}{\mathrm{d}E} =
\frac{(2.3\pm0.6_\text{stat}\pm0.2_\text{syst}) \cdot 10^{-13}}
{\mathrm{TeV}\,\mathrm{cm}^2\,\mathrm{s}} \frac{E}{1\,
\mathrm{TeV}}^{-2.5\pm0.5_\text{stat}\pm0.2_\text{syst}} \, ,
\end{equation*}
with a $\chi^2/{\mathrm{dof}} = 0.42/2$. The spectral slope is
compatible within errors with those previously reported by MAGIC
~\citep{MAGIC_lsi_science:2006vk,MAGIC_lsi_periodic:2009ApJ...693..303A,MAGIC_lsi_xrayvhe:2009ApJ...706L..27A}.
Hence, no evidence for long term spectral variability despite very
different fluxes during these different campaigns is observed.
Moreover, a exponential cutoff was also fitted allowing the power
law parameters to vary in the one sigma range (adding linearly
statistical and systematic uncertainties) with respect to the
fitted spectra in~\cite{MAGIC_lsi_xrayvhe:2009ApJ...706L..27A}.
That leads to a best fit cutoff at 483 GeV with a reduced
$\chi^2/\mathrm{dof} = 4.8 / 1$, which is strongly increased with
respect to the power law fit and hence strongly disfavors a cutoff
in the spectra to explain the reduced flux level.

\section{Discussion}

The binary system \lsi\ was detected emitting VHE $\gamma$-rays in
2009--2010 at a level a factor 10 lower than previously observed in the
phase interval 0.6--0.7. The previously observed orbital modulated
outburst in this interval was not detected during the first
observed orbital cycle, whereas it was observed in the second.
From this data set alone it was not possible to determine whether
the outburst is still a truly periodic feature of the light curve.
There are also other orbital phases, varying from one orbit to
another, during which significant emission was observed from 2005
to 2008. These individual nights were rarely observed and
contributed only a minor fraction to the integral signal from
\lsi\ in these campaigns. Although we could not significantly
detect emission at individual orbital phases in the here presented
observations because of the weakness of the source, it appears
that several phase intervals dominate in the integral signal. This
is strong evidence for a new behavior in the VHE $\gamma$-ray
emission of \lsi. In previous observations conducted with MAGIC,
the 0.6--0.7 interval dominated the total flux.

Furthermore it is evident that the flux during that phase interval
is considerable reduced compared to the previous campaigns and on
a similar level as in other phase intervals (e.g. 0.9--1.0). This
suggests that a change in the VHE $\gamma$-ray emission of \lsi\
has occurred. On the other hand, there was no statistically
significant  change in the spectrum of the orbit-integrated flux
in 2009 compared to the earlier results, suggesting that  the same
processes continue to produce VHE gamma-rays, but that either
fewer are produced or they are more absorbed.

If enhanced opacity causes the observed decreases in the VHE flux,
the photons will be redistributed to lower energies and thus might
be visible as a flux enhancement. Those VHE $\gamma$-rays would
thus need to propagate through a circum-source environment with a
higher photon energy density and we would expect to detect a
cut-off or an absorption feature in our spectrum. A cut-off in our
energy range that reduces the integral flux by a factor of ten
compared to our previous measurements
(e.g.~\cite{MAGIC_lsi_xrayvhe:2009ApJ...706L..27A})is incompatible
with our spectrum. The spectral fit disfavors an increased
photon-photon absorption around the emission region as the
explanation for the flux reduction in the VHE domain.

Another possibility is that there might be fewer accelerated
particles and/or fewer seed photons or less target matter,
resulting in lower VHE $\gamma$-rays production, depending on the
details of the assumed model (microquasar, pulsar wind, leptonic
or hadronic production). However, regardless of the scenario, a
change in the stellar wind density profile might explain the
change in the VHE $\gamma$-ray emission level: the wind density,
velocity, and porosity determine the accretion rate in the
microquasar scenario and the location of the termination shock in
the pulsar wind scenario. Depending on the magnitude of these
changes it might be difficult to explain the large variation in
the VHE domain, at least a recent study of the effect of wind
clumping in the framework of a microquasar scenario found only
variations of about 10\%~\citep{Owocki_2009}. In addition, the
effect of possible stellar wind density variations on the VHE
emission in \lsi\ are not yet well understood.

The VERITAS observations in the same period as we have presented
here did not detect VHE $\gamma$-ray emission from the
system~\citep{VERITAS_lsi_2011arXiv1105.0449A}. Our measurements
are not, however, in contradiction to those of VERITAS. Our longer
integration combined with a denser sampling of two orbital cycles
yielded a fainter detection threshold than from previous campaigns
expected VHE $\gamma$-ray signal from \lsi. Thus it is evident
that a frequent sampling with long individual integrations is
required not to miss weak emission from binary systems.

This is the first VHE $\gamma$-ray detection of \lsi\ in the era
of the \emph{Fermi} satellite. The faint emission at VHE
$\gamma$-rays does not yet permit night by night correlation
studies but do show that the emission in \lsi\ has changed on
longer timescale, since 2007. More sensitive and even deeper VHE
$\gamma$-ray observations should yield shorter timescale
correlation studies.

\section*{Acknowledgments}

We would like to thank the Instituto de Astrof\'{\i}sica de
Canarias for the excellent working conditions at the Observatorio
del Roque de los Muchachos in La Palma. The support of the German
BMBF and MPG, the Italian INFN, the Swiss National Fund SNF, and
the Spanish MICINN is gratefully acknowledged. This work was also
supported by the Marie Curie program, by the CPAN CSD2007-00042
and MultiDark CSD2009-00064 projects of the Spanish
Consolider-Ingenio 2010 programme, by grant DO02-353 of the
Bulgarian NSF, by grant 127740 of the Academy of Finland, by the
YIP of the Helmholtz Gemeinschaft, by the DFG Cluster of
Excellence ``Origin and Structure of the Universe'', and by the
Polish MNiSzW Grant N N203 390834.

Facilities: \facility{MAGIC}


\bibliographystyle{astron}

\end{document}